\documentclass[prl,aps,superscriptaddress,onecolumn,notitlepage,showpacs]{revtex4-1}
\usepackage{latexsym}
\usepackage{amsmath}
\usepackage{amssymb}
\usepackage{graphicx}
\usepackage{caption}
\usepackage{subfigure}
\usepackage{float}
\usepackage{mathrsfs}
\usepackage{color}
\usepackage{txfonts}
\usepackage[justification=centering,
            format=plain]{caption}

\renewcommand{\raggedright}{\leftskip=0pt \rightskip=0pt plus 0cm}

\begin{document}

\title{Local fluctuations of vibrational polaritons monitored by two-dimensional infrared spectroscopy}

\author{Zhedong Zhang}
\email{zhedong.zhang@tamu.edu}
\affiliation{Institute for Quantum Science and Engineering, Texas A$\&$M University, College Station, TX 77843, USA}

\author{Kai Wang}
\affiliation{Institute for Quantum Science and Engineering, Texas A$\&$M University, College Station, TX 77843, USA}

\author{Zhenhuan Yi}
\affiliation{Institute for Quantum Science and Engineering, Texas A$\&$M University, College Station, TX 77843, USA}

\author{Shaul Mukamel}
\affiliation{Department of Chemistry, University of California Irvine, Irvine, CA 92697, USA}

\author{Marlan O. Scully}
\affiliation{Institute for Quantum Science and Engineering, Texas A$\&$M University, College Station, TX 77843, USA}
\affiliation{Quantum Optics Laboratory, Baylor Research and Innovation Collaborative, Waco, TX 76704, USA}
\affiliation{Department of Mechanical and Aerospace Engineering, Princeton University, Princeton, NJ 08544, USA}

\date{\today}

\begin{abstract}
We study the collective behavior of molecules placed in an infrared (IR) microcavity, incorporating the local fluctuations, i.e., dynamical disorder. The cooperative feature in vibrational polaritons is shown to be dynamically eroded, due to intermolecule coherence. To further resolve such process, we develop a two-dimensional infrared spectroscopy (2D-IR) for molecules interacting with cavity modes. 
The cooperative feature in correspondence to the spectroscopic signal is specified. The results reveal the dark states by the cross peaks apart from the ones for polaritons, as a result of the breakdown of cooperativity between molecules. We further show that the breakdown of cooperativity profoundly connects to the localization of the vibrational excitations whereas the polariton modes are extended wave over several molecules. Besides, our work offers new physical insight for understanding the recent 2D-IR experiments where the interaction between dark modes and bright polaritons was evident.
\end{abstract}

\maketitle

\section{Introduction}
Microcavities open up a new way to access the strong coupling regime between material and photons \cite{Nuszmann_NatPhys2005,Forrest_NatPht2010,Cacciola_ACSNano2014,Kondo_APL2014}. The underlying theoretical framework is the cavity quantum electrodynamics (Cavity-QED), which has been well developed for atomic ensembles over decades \cite{Haroche_PT1989,Haroche_RMP2001,Scully_PRL2003,Scully_PNAS2011}. The hybridization of material excitations  with photons leads to a joint matter-photon states as referred to polaritons, which enables new optical properties with a wide range of application, such as polariton condensation in semiconductors \cite{Kasprzak_Nat2006,Balili_Sci2007}, modification of energy transfer pathways in organic molecules \cite{Zhang_SR2016,Zhang_JPCB2015,Coles_NatMat2014,Coles_NatCommun2014,Zhang_JCP2018} and the manipulation of chemical kinetics \cite{Mukamel_JPCL2016,Bellessa_PRL2004,Ebbesen_NatCommun2015}.

The interaction between matter and vacuum photon mode results in the two polariton branches, which are separated by the Rabi splitting energy $\hbar\Omega_R$ \cite{Haroche_RMP2001}. In samples containing number of molecules, i.e., J- and H-aggregate, this Rabi splitting scales as $\sqrt{N/V}$ where $N$ and $V$ are the amount of molecules and cavity volume, respectively. The volume $V$ of a microcavity is typically very large $\sim \mu\text{m}^3$ compared with the scale of a single molecule $\sim\text{nm}^3$ \cite{Ebbesen_ACR2016}. This offers the opportunity to couple several molecules to a single-mode cavity and the Rabi splitting will be then considerably enhanced. The strong coupling of molecules to cavity photon modes was reported recently \cite{Ebbesen_JPCL2015,Muallem_JPCL2016,Ebbesen_AC2013}. Organic molecules present a particularly favor case, due to the large dipole moments resulting in the Rabi splitting up to 1eV, a considerable fraction of the molecular transition energy \cite{Lidzey_Nat1998,Ebbesen_PRL2011}. These achievements inspire further investigations of the role of nuclei motion which could break the Born-Oppenheimer approximation \cite{Feist_PRX2015,Kowalewski_JCP2016}, since it is still an open issue in polariton dynamics. Even for the ground state, the chemical reactivity is considerably modified by showing the suppression of reaction rate when the Si-C vibrational stretching modes of reactant were strongly coupled to infrared microcavities \cite{Ebbesen_ACIE2016,Ebbesen_AC2012}. The photoluminescence spectroscopy was employed to demonstrate the elimination of vibronic coupling in J-aggregates \cite{Spano_JCP2015,Zhang_CPL2017}. Despite all these developments, the dynamics of many molecules in response to nuclear-induced fluctuations still remains elusive, especially when the strong coupling to photon modes presents.

Cooperativity in atomic ensembles is one of the most important topics in cavity QED \cite{Dicke_PR1954,Cummings_PR1968,Chumakov_QSO1996,Bauer_PRL2013}. It is reflected by the $\sqrt{N}$-scaling of Rabi splitting between polariton branches. Such collective nature stems from the photon-mediated interaction, resulting in the correlation between atoms. When it comes to molecular systems, the situation becomes complicated and obscured because of the entanglement between different degrees of freedoms even in single molecules \cite{Zhang_JPCL2017}. For instance, the nuclear motion that causes the exciton dephasing apparently plays an important role in understanding the exciton relaxation in molecules \cite{Ebbesen_NatCommun2015,Ebbesen_JPCL2015,Muallem_JPCL2016,Lidzey_Nat1998}. 
Recent study on CO bond stretching of polyvinyl acetate manifests the role of dark states in the vibrational relaxation under the influence of low-energy rovibrational modes \cite{Vidal_NJP2015}. The fluctuations produced by low-frequency nuclear modes can destroy the intermolecule coherence. This would affect molecule-photon interaction and break the cooperativity between molecules and further modify the chemical reaction kinetics.


In this article we address the issue of local fluctuation effect on the dynamics of the collective excitations in molecular ensembles. To this end, we develop a third-order resonant IR spectroscopy for a sample containing many molecules in an IR microcavity, incorporating the disorder effect as typical local fluctuations. We will demonstrate how the dynamical disorder erodes the cooperativity between molecules and subsequently lead to the dark states in weak coupling to the cavity modes. The sample is pumped by three time-ordered pulses 
and the photon-echoes signal is collected by heterodyne detection. The variation of $T_2$ delay provides the information regarding the relaxation of vibrational polaritons. We demonstrate the effect of disorder characterized by the cross peaks corresponding to the dark modes. This enables us to gain new insight for understanding the recent experiments which demonstrated the coupling between dark modes and bright polaritons \cite{Xiang_PNAS2018}. Besides, we find that the localization of vibrational excitations owing to the inter-molecule coherence (quantum interference) plays a significant role in understanding these features.

\begin{figure}
 \captionsetup{justification=raggedright,singlelinecheck=false}
 \centering
   \includegraphics[scale=0.25]{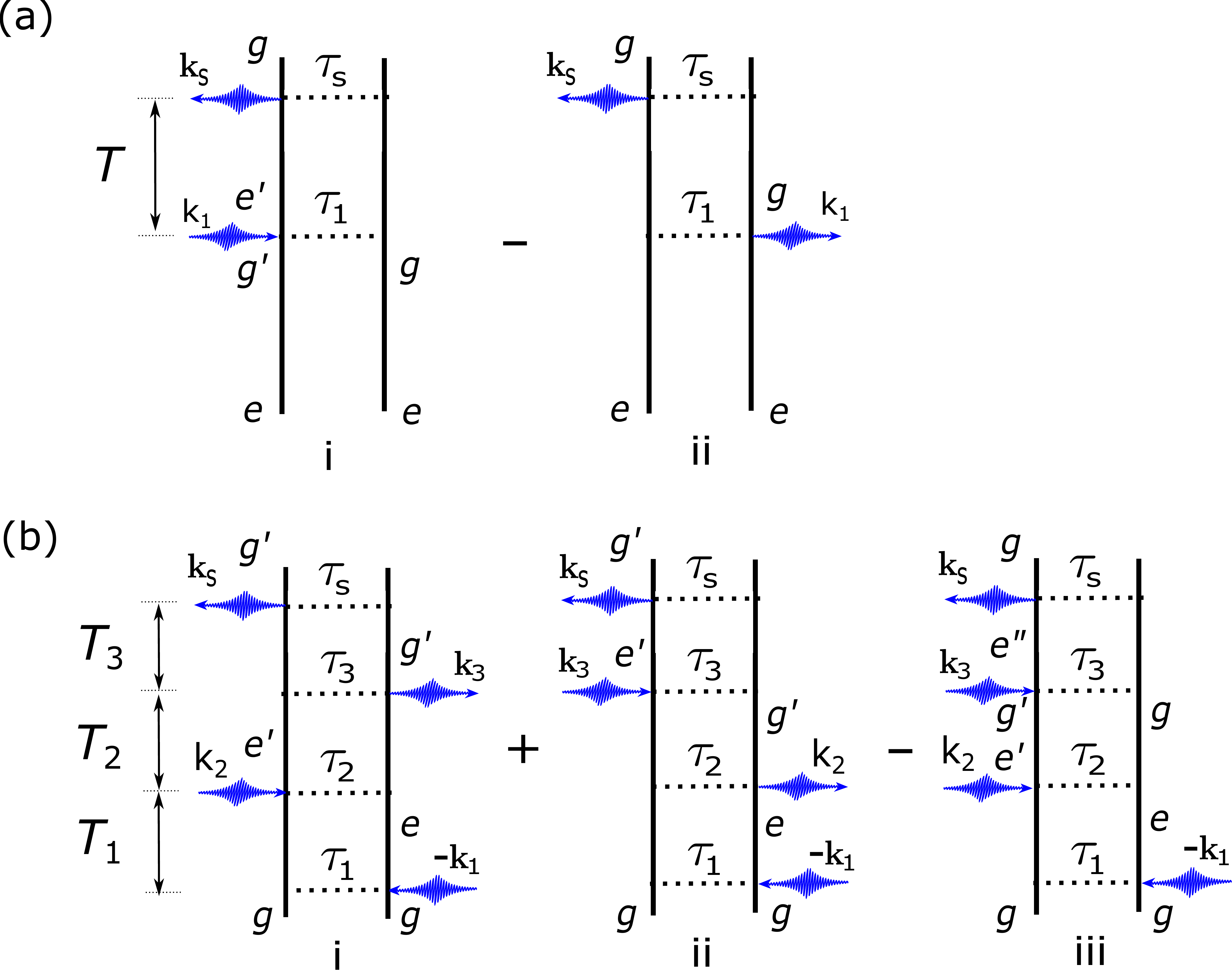}
\caption{Double-side Feynman diagrams for (a) TRPS and (b) photon-echo signal in 2D-IR spectra. (b) Processes (i), (ii) and (iii) correspond to excited-state emission (ESE), ground-state bleaching (GSB) and excited-state decay (ESD), respectively. The (i) in (a) and (iii) in (b) are due to cavity leakage.}
\label{fd}
\end{figure}

\section{Model and Equation of Motion}

\begin{figure}
 \captionsetup{justification=raggedright,singlelinecheck=false}
 \centering
   \includegraphics[scale=0.42]{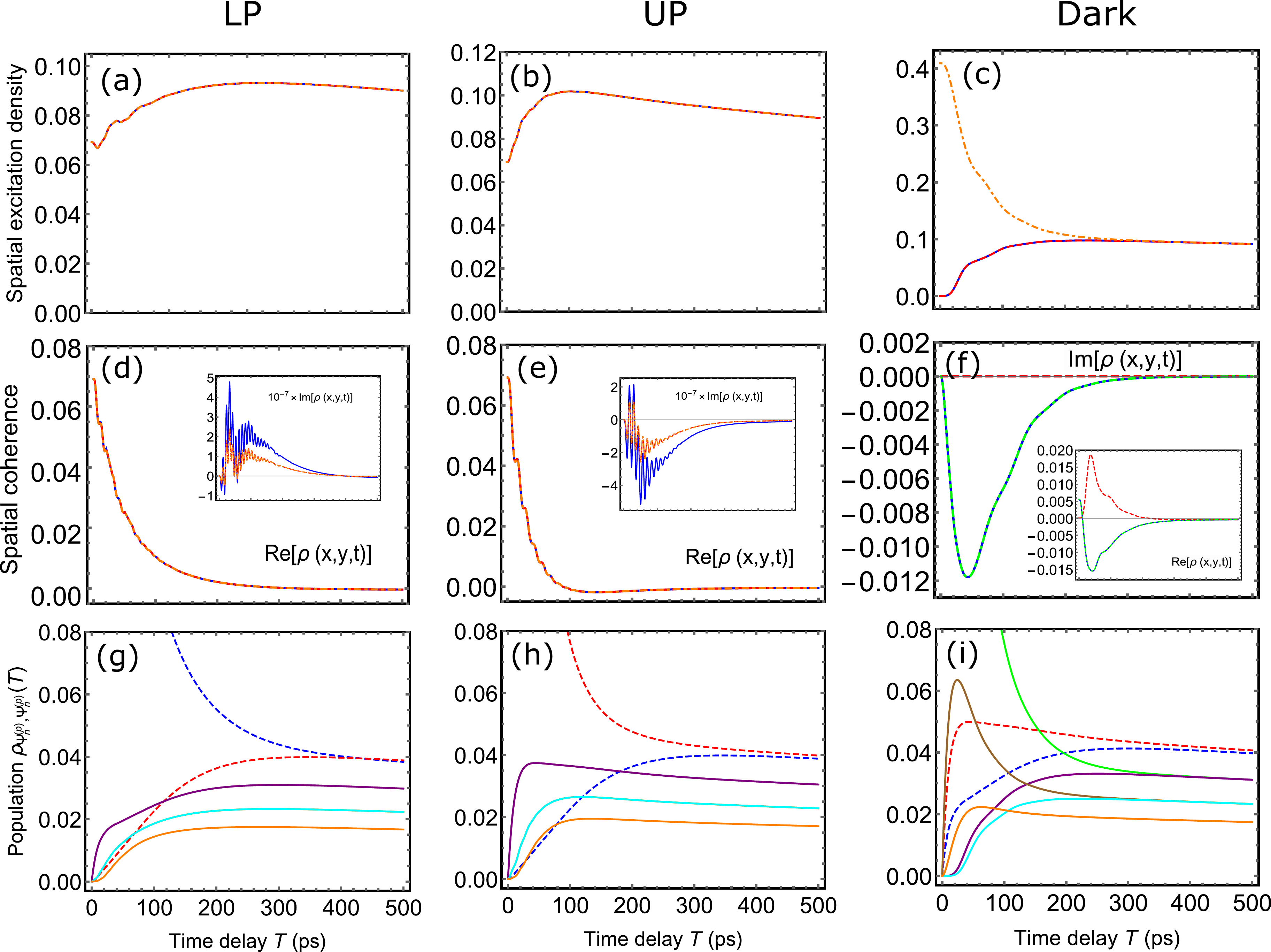}
\caption{Time-evolution of density matrix of three W(CO)$_6$ molecules placed in IR cavity with a linear fashion along the cavity axis. (Top) Population of the excitations at each molecule; (Middle) Intermolecule coherence; (Bottom) Polariton populations. In bottom row, dashed blue and dashed red lines correspond to LP and UP, respectively; Solid lines in bottom row correspond to dark states. Molecular parameters are $\omega_j=1983$cm$^{-1}$, $\omega_c=1983$cm$^{-1}$, $\delta\omega_j=18$cm$^{-1}$, $g_j=2.1$cm$^{-1}$, $v_j=62$cm$^{-1}$, $\gamma_j=0.18$cm$^{-1}$, $\omega_c/Q=0.04$cm$^{-1}$ and $T=300$K.}
\label{dm}
\end{figure}

Let us consider a sample containing a group of molecules where the surrounding environment (i.e., solvent) causes local fluctuations of the vibrational frequencies that is responsible for the disorder. This is quantified by an extra term $\Delta\omega(\{q_l\})$ in addition to the frequency of molecular vibration $\omega$ and $\{q_l\}$ denotes the collective coordinates of solvent. Usually the coordinate $q_l$'s are stochastic variables due to the large amount of low-frequency modes in solvent whose coupling to the vibrational modes is ignorable. For simplicity, we take into account of a single coordinate here, namely, $\Delta\omega(\{q_l\})=\Delta\omega(q)$. Since we are aiming to understand the underlying physics of the local fluctuation effect on vibrational polaritons, we can further adopt the two-state description for solvent motion, namely, $q=0,1$. This recasts the discrete quantum jump model where the low-frequency modes acting as a thermal bath with a smooth spectral density leads to the random transition between these states. In the rotating frame of photon, such hybrid system is described by the Hamiltonians
\begin{equation}
\begin{split}
& H_p = \hbar\sum_{i=1}^N\left[\left(\omega_i + \delta\omega_i\eta_i^z - \omega_c\right)b_i^{\dagger}b_i + \Delta_i b_i^{\dagger}b_i^{\dagger}b_i b_i\right] + \hbar\sum_{i=1}^N g_i\left(b_i^{\dagger}a + b_i a^{\dagger}\right)\\[0.15cm]
& V_{pl-env}(t) = \hbar\sum_{i=1}^N\sum_s\lambda_{i,s}\left(\sigma_i^+B_s^{(i)}e^{i(v_i-v_s^{(i)})t} + \sigma_i^-B_s^{(i),\dagger}e^{-i(v_i-v_s^{(i)})t}\right),\ \ H_{env} = \hbar\sum_{i=1}^N\sum_s v_s^{(i)}B_s^{(i),\dagger}B_s^{(i)}
\end{split}
\label{ht}
\end{equation}
where $b_j,\ a$ and $B_s^{(j)}$ are the bosonic annihilation operators for the vibration of the $j$-th molecule, cavity photon and thermal bath, respectively. $\eta_j^z=\frac{1}{2}(1-\sigma_j^z)$ and $\sigma_j^z,\sigma_j^{\pm}$ represent the Pauli matrices operating on solvent coordinate at the $j$-th molecule: $\sigma_j^+|q_j\rangle=\sqrt{1-q_j}\ |1-q_j\rangle,\ \sigma_j^-|q_j\rangle=\sqrt{q_j}\ |1-q_j\rangle,\ \sigma_j^z|q_j\rangle=(1-2q_j)|q_j\rangle$. $\omega_j$ and $\Delta_j$ stand for the frequency and anharmonic interaction of vibrational mode of the $j$-th molecule. $\omega_c$ is the photon frequency and $v_i$ denotes the solvent energy difference between $q_i=0,1$, located at the $i$-th molecule. The weak system-bath coupling and Markovian approximations give rise to the Quantum Master Equation (QME)
\begin{equation}
\begin{split}
\dot{\rho} = & \frac{i}{\hbar}[\rho,H_p] + \sum_{i=1}^N\frac{\gamma_i}{2}\Big[\bar{n}_{v_i}\left(2\sigma_i^+\rho\sigma_i^- - \sigma_i^-\sigma_i^+\rho -\rho\sigma_i^-\sigma_i^+\right)\\[0.15cm]
& \qquad + (\bar{n}_{v_i}+1)\left(2\sigma_i^-\rho\sigma_i^+ - \sigma_i^+\sigma_i^-\rho -\rho\sigma_i^+\sigma_i^-\right)\Big] + \frac{\omega_c}{2Q}\left(2a\rho a^{\dagger} - a^{\dagger}a\rho - \rho a^{\dagger}a\right)
\end{split}
\label{qme}
\end{equation}
which recasts the Quantum Stochastic Liouville Equation (QSLE) formulated at phenomenological level before \cite{Tanimura_JPSJ2006,Mukamel_ChemRev2009}. $\gamma_i=2\pi\sum_s\lambda_{i,s}^2\delta(v_i-v_s^{(i)})$ and $Q$ denotes the quality factor of the infrared cavity. As $[H_p,\sigma_j^z]=0$, the Hamiltonian $H_p$ is block diagonal
\begin{equation}
\begin{split}
H_p = \bigoplus\sum_{\{l_1l_2\cdots l_N\}}H_p^{(l_1l_2\cdots l_N)}
\end{split}
\label{hp}
\end{equation}
under the basis $|n_1,n_2,\cdots,n_N;m\rangle\otimes|l_1,l_2,\cdots,l_N\rangle$ where $n_j$ and $m$ denote the numbers of vibrational excitations on the $j$-th molecule and photons, respectively. $l_1,l_2,\cdots,l_N$ denotes the configurations of the solvent coordinates and $l_j=0,1;j=1,2,\cdots,N$. Since the total excitation number $M=\sum_{i=1}^N n_i+m$ is conserved, we only consider the ground-state and single-excited state manifolds, namely, $M=0,1$, whose basis are $|0_1,0_2,\cdots,0_N;0\rangle\otimes|l_1,l_2,\cdots,l_N\rangle$ and $|e_i\rangle\otimes|l_1,l_2,\cdots,l_N\rangle=\{|0_1,0_2,\cdots,1_j,\cdots,0_N;0\rangle\otimes|l_1,l_2,\cdots,l_N\rangle;\ |0_1,0_2,\cdots,0_N;1\rangle\otimes|l_1,l_2,\cdots,l_N\rangle\}$. Let $|\psi_k^{(l_1l_2\cdots l_N)}\rangle$ be the $k$-th eigenstate of $H_p^{(l_1l_2\cdots l_N)}$ for $M=1$ manifold so that
\begin{equation}
\begin{split}
|\psi_k^{(l_1l_2\cdots l_N)}\rangle = \sum_{j=1}^{N+1}C_{j,k}^{(l_1l_2\cdots l_N)}|e_j\rangle\otimes|l_1,l_2,\cdots,l_N\rangle
\end{split}
\label{C}
\end{equation}
where $C^{(l_1l_2\cdots l_N)}$ is the unitary matrix diagonalizing $H_p^{(l_1l_2\cdots l_N)}$. Under the resonant condition $\omega_c\simeq \omega_j$, $H_p^{(0_1,0_2,\cdots,0_N)}$ gives rise to two Dicke states which are referred to lower polariton (LP) and upper polairton (UP). The eigenstate for $M=0$ manifold is $|G^{(l_1l_2\cdots l_N)}\rangle=|0_1,0_2,\cdots,0_N;0\rangle\otimes|l_1,l_2,\cdots,l_N\rangle$. Introducing the bookkeeping notation $P,Y$ for denoting the configurations $\{l_1,l_2,\cdots,l_N\},\{r_1,r_2,\cdots,r_N\}$, some algebra gives the following equations of motion
\begin{equation}
\begin{split}
& \frac{\text{d}}{\text{d}t}\langle\langle e_{n'}e_n;P|\rho\rangle\rangle = \sum_{m',m=1}^{N+1}\sum_Y\langle\langle e_{n'}e_n;P|\hat{L}|e_{m'}e_m;Y\rangle\rangle\langle\langle e_{m'}e_m;Y|\rho\rangle\rangle\\[0.15cm]
& \frac{\text{d}}{\text{d}t}\langle\langle G^{(P)},G^{(P)}|\rho\rangle\rangle = \sum_Y \langle\langle G^{(P)},G^{(P)}|\hat{L}|G^{(Y)},G^{(Y)}\rangle\rangle \langle\langle G^{(Y)},G^{(Y)}|\rho\rangle\rangle + \frac{\omega_c}{Q}\langle\langle e_{N+1}e_{N+1};P|\rho\rangle\rangle\\[0.15cm]
& \frac{\text{d}}{\text{d}t}\langle\langle G^{(P)},\psi_j^{(P)}|\rho\rangle\rangle = \left(i\omega_j^{(P)} - \sum_{m=1}^N\gamma_m\left(\bar{n}_{v_m}+l_m\right)\right)\langle\langle G^{(P)},\psi_j^{(P)}|\rho\rangle\rangle
\end{split}
\label{rho}
\end{equation}
where the matrix elements $\langle\langle e_{n'}e_n;P|\hat{L}|e_{m'}e_m;Y\rangle\rangle$ and $\langle\langle G^{(P)},G^{(P)}|\hat{L}|G^{(Y)},G^{(Y)}\rangle\rangle$ are determined by Eq.(\ref{qme}). $|e_{n'}e_n;P\rangle\rangle\equiv |e_{n'}e_n\rangle\rangle\otimes|P,P\rangle\rangle$. This set of equations dictate the coupling between polariton and dark states under the dynamical disorder which erodes the collective nature of the vibrational polaritons. To elucidate this, let us proceed via the solution to the first line in Eq.(\ref{rho})
\begin{equation}
\begin{split}
|\rho(t)\rangle\rangle = \sum_{i,j=1}^{N+1}\sum_{k,l=1}^{N+1}\sum_P\sum_{u=1}^{\text{dim}(L)}S_{\langle e_ie_j,P\rangle;u}e^{\nu_u t}S_{u;\langle e_ke_l,J\rangle}^{-1} C_{e_k,a}^{(J),*}C_{e_l,a}^{(J)}|e_ie_j;P\rangle\rangle
\end{split}
\label{rhosol}
\end{equation}
with the initial condition $|\rho(0)\rangle\rangle=|\psi_a^{(J)},\psi_a^{(J)}\rangle\rangle$. $\nu_n$'s are the eigenvalues of Liouvillian $\hat{L}$ with negative real part and $\text{dim}(L)=(N+1)^2 2^N$. Then in coordinate space one has $\rho(x,x';t)=\langle\langle x,x'|\rho(t)\rangle\rangle$, yielding to
\begin{equation}
\begin{split}
\rho(x,x';t) = \sum_{i,j=1}^{N+1}\sum_{k,l=1}^{N+1}\sum_P\sum_{u=1}^{\text{dim}(L)}S_{\langle e_ie_j,P\rangle;u}e^{\nu_u t}S_{u;\langle e_ke_l,J\rangle}^{-1} C_{e_k,a}^{(J),*}C_{e_l,a}^{(J)}\ \varphi(x-a_i)\varphi^*(x'-a_j)
\end{split}
\label{rhox}
\end{equation}
where $\varphi(x-a_i)=\sqrt{\frac{2}{\ell_i^3\sqrt{\pi}}}\ (x-a_i)e^{-(x-a_i)^2/2\ell_i^2}$ is the wave function of single vibrational excitation at the $i$-th molecule. The $i$-th molecule locates at position $a_i$ and $\ell_i$ defines the typical length of vibrations at the $i$-th molecule. $x'=x$ gives the spatial density of vibrational excitations while $x'\neq x$ gives the intermolecule quantum coherence. Because of the negative $\text{Re}(\nu_n)$, both density and coherence will decay as time propagates. However, the coherence $\rho(x,x',t)$ decays much faster than the density $\rho(x,x,t)$. This is demonstrated by comparing the top and middle rows of Fig.\ref{dm}. In these simulations we consider three W$(\text{CO})_6$ molecules placed in an IR-cavity where the joint-vibration/photon system is initially prepared at either polariton state (LP or UP), under strong disorder such that the vibration-photon interaction $g_j\ (j=1,2,\cdots,N)$ is weaker than the fluctuation of vibrational frequencies, namely, $g_j<\delta\omega_j$. This results in the excitation of the joint vibration/photon system localized in the vicinity of the position that individual molecule places whereas the polariton modes are extended wave over whole ensemble. These localized waves are not correlated with each other, as evident by the fast decay of intermolecule coherence. Thereby each localized excitation is the  consequence of a certain coherent superposition of several eigenmodes of the hybrid system that extend over the whole ensemble. In this sense, the localization of vibrational excitations predicted by Eq.(\ref{rhox}) and Fig.\ref{dm} shows the analogy to the Anderson localization mechanism of both electrons and photons in disordered materials \cite{Anderson_PR1958,Agarwal_PRA2010,Wiersma_Nat1997}.

On the other hand, the loss of intermolecule coherence erodes the collective nature of the vibrational polaritons, resulting in the dark states which weakly interact with cavity photons. Thus the excitation transfer from polariton to these dark states shows up, as revealed by polariton dynamics in Fig.\ref{dm}(g) and \ref{dm}(h). Suppose the joint vibration/photon system is engineered at the dark states, we clearly observe in Fig.\ref{dm}(f) a rapid increase of intermolecule coherence during the first $\sim 40$ps and a rapid decay afterward. This implies the dark-states$\rightarrow$polariton transfer and a subsequent transfer of polariton$\rightarrow$dark-states again. The excitation transfer between polaritons and dark modes is actually faster than the one between polariton branches, as illustrated by population dynamics in Fig.\ref{dm}(g), \ref{dm}(h) and \ref{dm}(i). Such coupling between polaritons and dark states will be further manifested in both the time-resolved photoluminescence and 2D-infrared spectroscopies which we will develop later on.

\begin{figure}
 \captionsetup{justification=raggedright,singlelinecheck=false}
 \centering
   \includegraphics[scale=0.31]{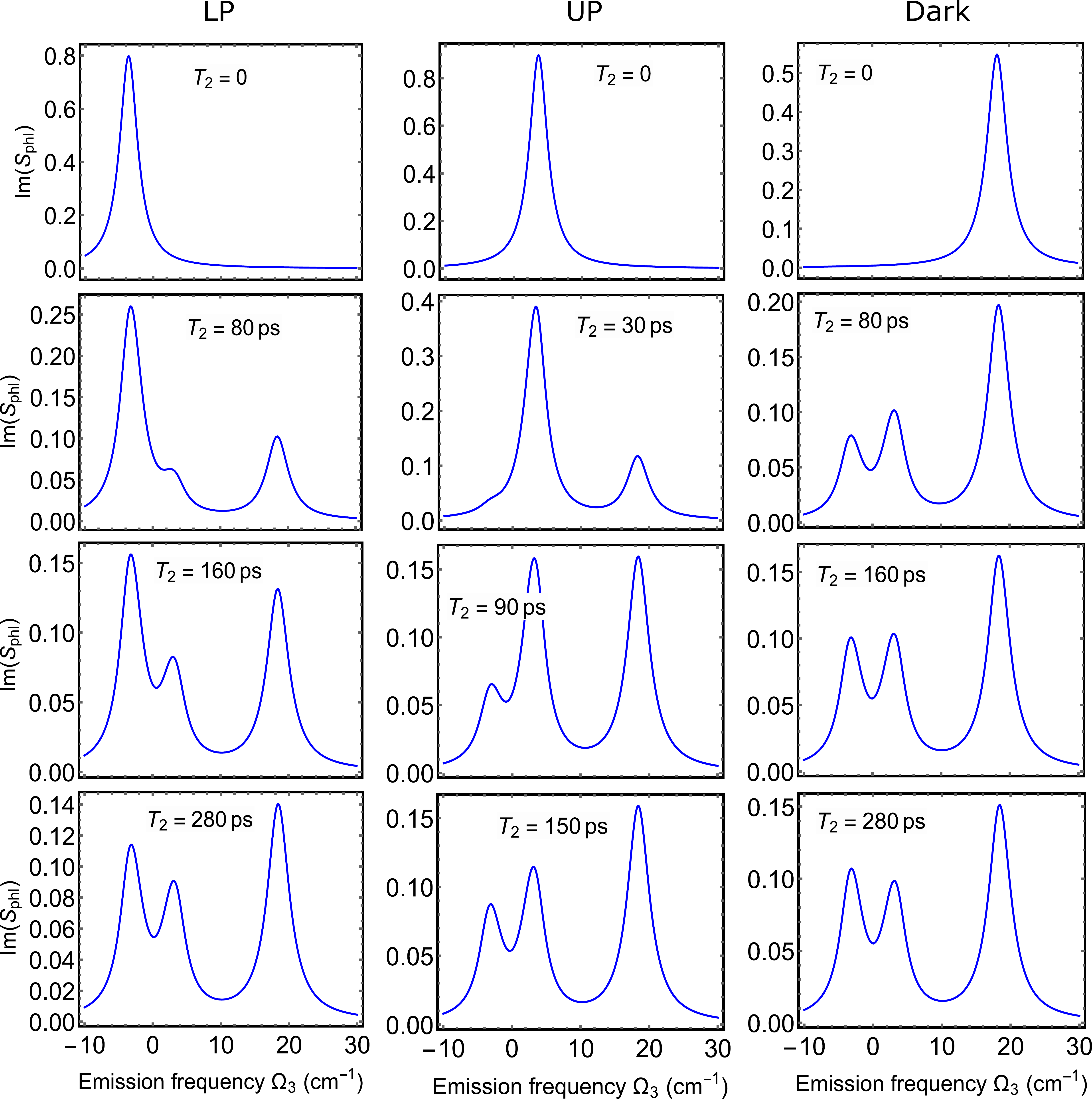}
\caption{TRPS varies with action time of probe pulse according to Eq.(\ref{Str}), for three W(CO)$_6$ molecules placed in an IR cavity with a linear fashion along the cavity axis. The joint vibration/photon system is prepared to (left column) LP, (middle column) UP and (right column) dark states; Pulse shape is set to be Gaussian; Electric dipole of CO-bond vibrations in W(CO)$_6$ molecule is $\mu=0.122$D. Molecular parameters are the same as Fig.\ref{dm}; Pulse parameters are $\sigma_{\text{pr}}=\sigma_{lo}=50$cm$^{-1}$ and $\omega_{\text{pr}}=\omega_{lo}=1993$cm$^{-1}$.}
\label{TRPS}
\end{figure}

\section{Time-resolved photoluminescence spectra for vibrational polaritons}
Suppose the sample is excited by actinic pulses, its relaxation could be probed by time-resolved photoluminescence spectroscopy (TRPS). The signal is collected by heterodyne detection where the local oscillator interferes with the radiated field after a time delay of $T$ with respect to the probe pulse, as depicted by Feynman diagram in Fig.\ref{fd}(a). Considering only the electric dipole $\boldsymbol{\mu}=\sum_{s=1}^N\boldsymbol{\mu}_s\left(b_s+b_s^{\dagger}\right)$, the interaction between the sample and probe field is of the dipolar form
\begin{equation}
\begin{split}
V_{int}(t) = \mu^{(\text{pr}),+}(t){\cal E}(t-\tau_{\text{pr}})e^{i(\textbf{k}_{\text{pr}}\cdot\textbf{r}-\omega_{\text{pr}}(t-\tau_{\text{pr}}))} + \text{h.c.}
\end{split}
\label{Vint}
\end{equation}
under rotating-wave approximation. $\mu^{(\text{pr}),+}(t)=\boldsymbol{\mu}^+(t)\cdot\textbf{e}_{\text{pr}}$ where $\boldsymbol{\mu}^+=\sum_{s=1}^N\boldsymbol{\mu}_s b_s^{\dagger}$ and $\textbf{e}_{\text{pr}}$ is the unit polarization vector of probe field. ${\cal E}(t-\tau)$ is the pulse envelop centered at time $\tau$. Thus the time-domain signal reads
\begin{equation}
\begin{split}
S(T,\tau_{\text{pr}}) = \left(-\frac{i}{\hbar}\right)(2\pi)^3\delta(\textbf{k}_{lo}-\textbf{k}_{\text{pr}})\int_{-\infty}^{\infty}\text{d}t\int_0^{\infty}\text{d}t_1\ \chi^{(1)}(t_1,t-t_1) E_{lo}^*(t-\tau_{lo})E_{\text{pr}}(t-t_1-\tau_{\text{pr}})
\end{split}
\label{Spht}
\end{equation}
with the first-order response function containing the dynamical information of sample
\begin{equation}
\begin{split}
\chi^{(1)}(t_1,t_2) = \langle\langle 1|\mu_L^{(lo),-}{\cal G}(t_1)\mu_L^{(\text{pr}),+}{\cal G}(t_2)|\psi_0,\psi_0\rangle\rangle - \langle\langle 1|\mu_L^{(lo),-}{\cal G}(t_1)\mu_R^{(\text{pr}),+}{\cal G}(t_2)|\psi_0,\psi_0\rangle\rangle
\end{split}
\label{x1}
\end{equation}
and ${\cal G}(t)$ stands for the free propagator in the absence of external fields. Notice that the 1st term in Eq.(\ref{x1}) originated from the cavity leakage is much smaller than the 2nd term, when using a good quality cavity with $\omega_c/Q\ll \gamma_i$. Such condition is necessary for observing the joint matter/photon states in strong coupling regime. 
In general the signal in Eq.(\ref{Spht}) is hard to calculate, due to the integrals over pulse shapes. But we will work under impulsive approximation \cite{Mukamel_book1995,Yan_JCP1991} where the time duration of pulse is short compared with the timescale of homogeneous dephasing as well as solvent reorganization processes. This is the case for many time-resolved spectroscopies. Some algebra gives the TRPS
\begin{equation}
\begin{split}
S(\Omega,\tau_{\text{pr}}) & = 2\text{Im}\int_0^{\infty} S(T,\tau_{\text{pr}})e^{i\Omega T}\text{d}T\\[0.15cm]
& = -\frac{16\pi^3}{\hbar}\delta(\textbf{k}_{lo}-\textbf{k}_{\text{pr}})\text{Im}\bigg[\bigg(\sum_Y\sum_{i,j=1}^{N+1} \frac{V_{(lo),i}^{(Y),*}V_{(\text{pr}),j}^{(Y)}}{\Omega-\omega_i^{(Y)}+i\gamma_i^{(Y)}} \langle\langle \psi_i^{(Y)},\psi_j^{(Y)}|\rho(\tau_{\text{pr}})\rangle\rangle\\[0.15cm]
& \qquad - \sum_Y\sum_{i=1}^{N+1} \frac{V_{(lo),i}^{(Y),*}V_{(\text{pr}),i}^{(Y)}}{\Omega-\omega_i^{(Y)}+i\gamma_i^{(Y)}}\langle\langle G^{(Y)},G^{(Y)}|\rho(\tau_{\text{pr}})\rangle\rangle\bigg)\ \tilde{{\cal E}}_{lo}^*(\omega_i^{(Y)}-\omega_{lo})\tilde{{\cal E}}_{\text{pr}}(\omega_i^{(Y)}-\omega_{\text{pr}})\bigg]
\end{split}
\label{Str}
\end{equation}
by Fourier transform with respect to $T$ delay, where $V_{(lo),i}^{(Y)}=\textbf{e}_{lo}\cdot\textbf{V}_i^{(Y)},\ V_{(\text{pr}),i}^{(Y)}=\textbf{e}_{\text{pr}}\cdot\textbf{V}_i^{(Y)}$. $\textbf{V}_i^{(Y)}=\sum_{s=1}^N\boldsymbol{\mu}_s C_{s,i}^{(Y)}\ (i=1,2,\cdots,N+1)$ are the matrix elements of dipole moment in eigenbasis of $H_p^{(Y)}$. In what follows we employ the Gaussian pulse shape $\tilde{{\cal E}}(\omega-v)={\cal E}_0 \text{exp}[-(\omega-v)^2/2\sigma^2]$ in the simulations, where $\sigma$ denotes the spectral width.

Fig.\ref{TRPS} shows the tomographies of TRPS with respect to different time delays of probe pulse. Starting from either polariton branch (LP or UP), we observe the extra peak at frequency $\simeq 18$cm$^{-1}$ without much shift as the time propagates, besides the peaks positioned at $\simeq \pm 3.6$cm$^{-1}$ with the separation of $\simeq 7.2$cm$^{-1}\simeq 2g\sqrt{N}$ corresponding to the two polariton branches. This indicates the dark states in weak interaction with cavity modes and also displays the excitation transfer between the bright polaritons and dark states. Moreover, by comparing the left and middle columns in Fig.\ref{TRPS}, one can clearly see the faster excitation transfer of UP $\rightarrow$ dark-states than that of LP $\rightarrow$ dark-states. This can be further understood by the faster decay of intermolecule coherence shown in Fig.\ref{dm}(d,e) when preparing the system at UP as well as the population dynamics shown in Fig.\ref{dm}(g,h). The right column in Fig.\ref{TRPS} illustrates the excitation transfer of dark-states $\rightarrow$ polaritons (LP and UP). In addition, the TRPS illustrates the quicker excitation transfer between polaritons and dark states than the one between LP and UP. This thereby offers the support to the previous conclusion from intermolecule coherence and population dynamics shown in Fig.\ref{dm}.

\begin{figure}
 \captionsetup{justification=raggedright,singlelinecheck=false}
 \centering
   \includegraphics[scale=0.4]{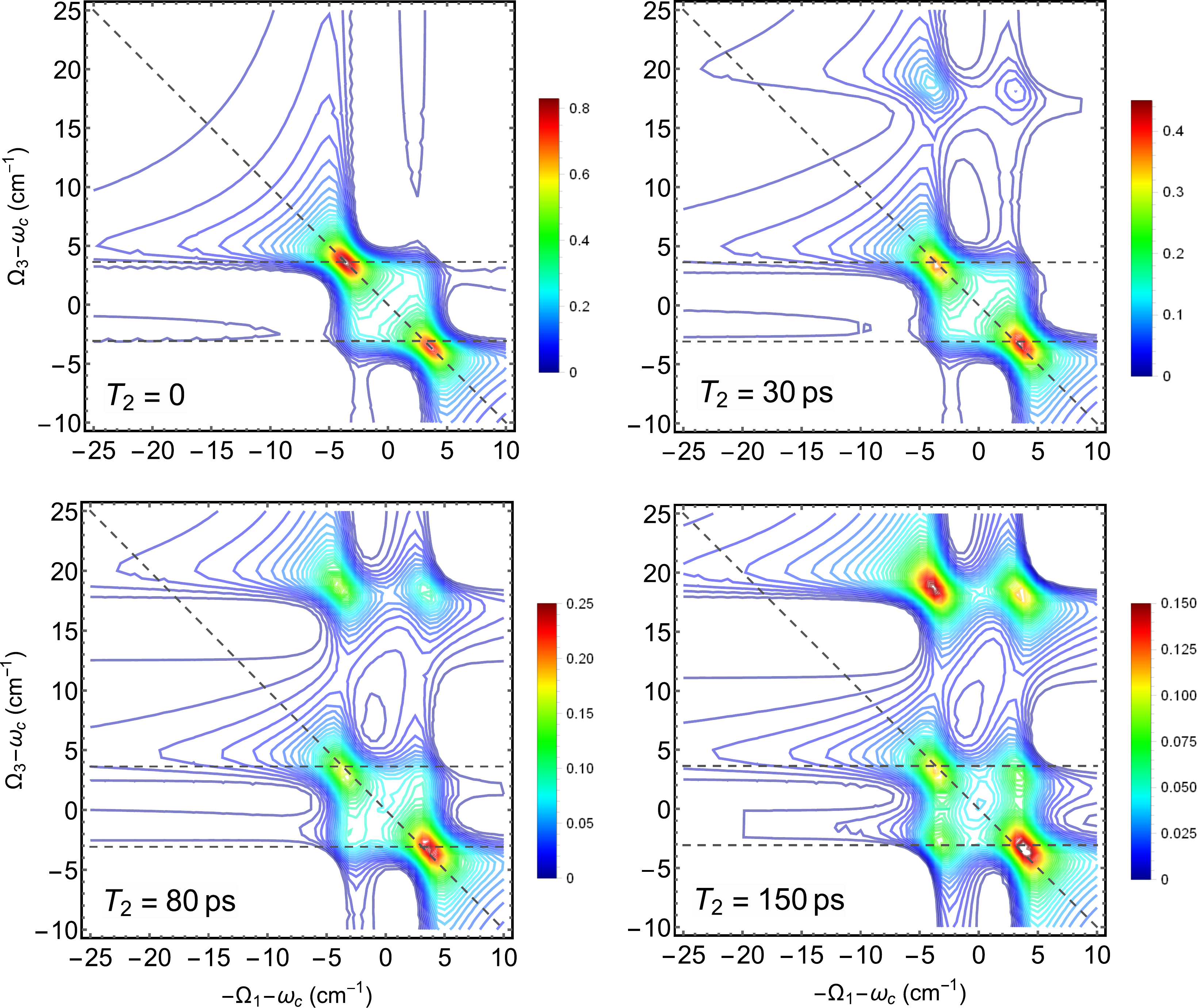}
\caption{2D-IR signal varies with $T_2$ delay according to Eq.(\ref{Sif}), for three W(CO)$_6$ molecules placed in an IR cavity with a linear fashion along the cavity axis. The joint vibration/photon system is at ground state where the solvent coordinate takes $q_j=0$, prior to pulse action; Molecular parameters are the same as Fig.\ref{dm}; Pulse parameters are $\sigma_1=\sigma_2=\sigma_3=\sigma_{lo}=50$cm$^{-1}$, $\omega_1=\omega_2=1983$cm$^{-1}$ and $\omega_3=\omega_{lo}=1993$cm$^{-1}$.}
\label{2DIR}
\end{figure}

\section{The third-order resonant infrared spectroscopy}

\begin{figure}
 \captionsetup{justification=raggedright,singlelinecheck=false}
 \centering
   \includegraphics[scale=0.4]{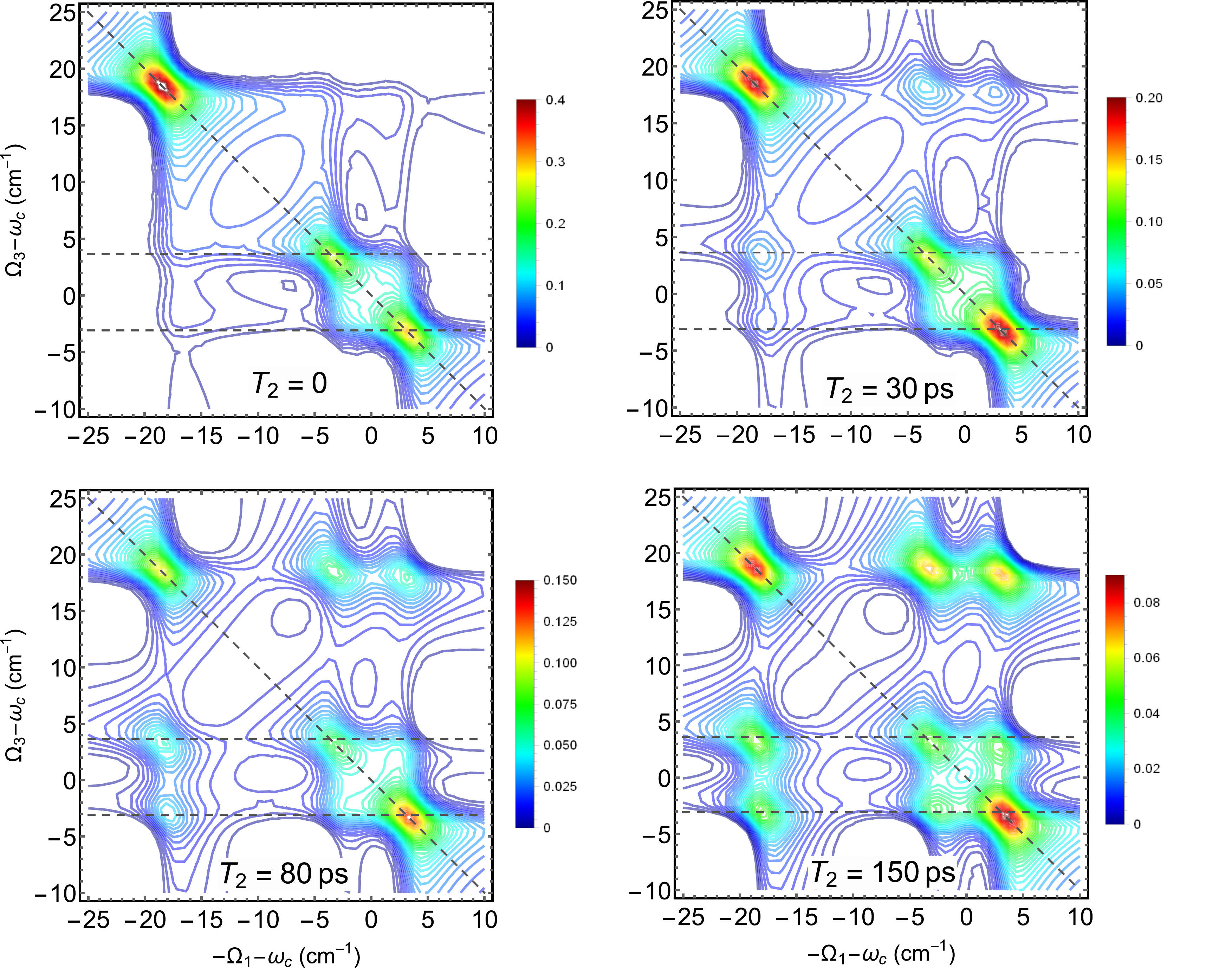}
\caption{2D-IR signal varies with $T_2$ delay according to Eq.(\ref{Sif}), for three W(CO)$_6$ molecules placed in an IR cavity with a linear fashion along the cavity axis. The joint vibration/photon system is at ground state where the solvent is at thermal equilibrium under room temperature, prior to the action of pulses; Pulse shape is set to be Gaussian. Molecular and pulse parameters are the same as Fig.\ref{2DIR}.}
\label{2DIRf}
\end{figure}

To gain more information regarding the relaxation of vibrational polaritons beyond the scope of TRPS, i.e., line broadening and transfer pathways, we will essentially develop a 2D-IR spectroscopy for the joint vibration/photon system by incorporating the disorder effect. The three related processes of excited-state emission (ESE), ground-state bleaching (GSB) and excited-state decay (ESD) are displayed in Fig.\ref{fd}(b). The sample interacts with three time-ordered pulses by means of dipolar coupling
\begin{equation}
\begin{split}
U_{int}(t) = \sum_{j=1}^3 \mu^{(j),+}(t){\cal E}(t-\tau_j)e^{i(\textbf{k}_j\cdot\textbf{r}-\omega_j(t-\tau_j))} + \text{h.c.}
\end{split}
\label{Uint}
\end{equation}
where $\mu^{(j),+}(t)=\boldsymbol{\mu}^+(t)\cdot\textbf{e}_j$ and $\textbf{e}_j$ is the unit polarization vector of the $j$-th pulse. The photon echoes field interferes in phase with the fourth pulse (local oscillator), and the signal is given by
\begin{equation}
\begin{split}
S_I(T_3,T_2,T_1) & = \int\text{d}^3\textbf{r}\int_{-\infty}^{\infty}\text{d}t\ \langle \mu^{(lo),-}\rangle_{\rho(t)}E_{lo}^*(t-\tau)e^{-i\textbf{k}_{lo}\cdot\textbf{r}}\\[0.15cm]
& = \frac{8\pi^3}{\hbar^3}\delta(\textbf{k}_{lo}-\textbf{k}_3-\textbf{k}_2+\textbf{k}_1)\int_{-\infty}^{\infty}\text{d}t \int_0^{\infty}\text{d}t_3 \int_0^{\infty}\text{d}t_2 \int_0^{\infty}\text{d}t_1\ \chi^{(3)}(t_3,t_2,t_1)\\[0.15cm]
& \qquad\qquad \times E_{lo}^*(t-\tau)E_3(t-t_3-\tau_3)E_2(t-t_3-t_2-\tau_2)E_1^*(t-t_3-t_2-t_1-\tau_1)
\end{split}
\label{Si}
\end{equation}
and the dynamical information of molecules is contained in the third-order response function
\begin{equation}
\begin{split}
\chi^{(3)}(t_3,t_2,t_1) = & \langle\langle 1|\mu_L^{(lo),-}{\cal G}(t_3)\mu_R^{(3),+}{\cal G}(t_2)\mu_L^{(2),+}{\cal G}(t_1)\mu_R^{(1),-}|G^{(J)},G^{(J)}\rangle\rangle\\[0.15cm]
& \quad + \langle\langle 1|\mu_L^{(lo),-}{\cal G}(t_3)\mu_L^{(3),+}{\cal G}(t_2)\mu_R^{(2),+}{\cal G}(t_1)\mu_R^{(1),-}|G^{(J)},G^{(J)}\rangle\rangle\\[0.15cm]
& \qquad - \langle\langle 1|\mu_L^{(lo),-}{\cal G}(t_3)\mu_L^{(3),+}{\cal G}(t_2)\mu_L^{(2),+}{\cal G}(t_1)\mu_R^{(1),-}|G^{(J)},G^{(J)}\rangle\rangle
\end{split}
\label{x3}
\end{equation}
where the 1st, 2nd and 3rd terms correspond to the ESE, GSB and ESD, respectively. In our model the 2nd term contains the information about the solvent relaxation itself, nothing to do with the vibrational polaritons. This contributes as a background which can be deduced from the full signal by running a control simulation. Under the impulsive approximation we insert Eq.(\ref{rho}) into $\chi^{(3)}(t_3,t_2,t_1)$ and carry out the multifold convolution with respect to pulse envelopes. Some algebra leads to the 2D signal
\begin{equation}
\begin{split}
S_I(\Omega_3,T_2,\Omega_1) & = -2\text{Im}\int_0^{\infty}\text{d}T_3 \int_0^{\infty}\text{d}T_1\ S_I(T_3,T_2,T_1)e^{i(\Omega_3T_3+\Omega_1T_1)}\\[0.15cm]
& = \frac{16\pi^3}{\hbar^3}\delta(\textbf{k}_{lo}-\textbf{k}_3-\textbf{k}_2+\textbf{k}_1)\text{Re}\bigg[\bigg(\sum_r\sum_{i,i'=1}^{N+1}\sum_J\sum_{j,j'=1}^{N+1}\frac{V_{(lo),i}^{(r),*}V_{(3),i'}^{(r)}V_{(2),j'}^{(J)}V_{(1),j}^{(J),*}}{(\Omega_3-\omega_i^{(r)}+i\gamma_i^{(r)})(\Omega_1+\omega_j^{(J)}+i\gamma_j^{(J)})}P_J\\[0.15cm]
& \qquad\ \times \langle\langle \psi_i^{(r)},\psi_{i'}^{(r)}|{\cal G}(T_2)|\psi_{j'}^{(J)},\psi_j^{(J)}\rangle\rangle + \sum_r\sum_{i=1}^{N+1}\sum_J\sum_{j=1}^{N+1}\frac{V_{(lo),i}^{(r),*}V_{(3),i}^{(r)}V_{(2),j}^{(J)}V_{(1),j}^{(J),*}}{(\Omega_3-\omega_i^{(r)}+i\gamma_i^{(r)})(\Omega_1+\omega_j^{(J)}+i\gamma_j^{(J)})}P_J\\[0.15cm]
& \qquad\quad \times \prod_{s=1}^N\langle\langle r_s,r_s|{\cal G}_s^{(gg)}(T_2)|J_s,J_s\rangle\rangle - \sum_r\sum_{i=1}^{N+1}\sum_J\sum_{j,j'=1}^{N+1}\frac{V_{(lo),i}^{(r),*}V_{(3),i}^{(r)}V_{(2),j'}^{(J)}V_{(1),j}^{(J),*}}{(\Omega_3 - \omega_i^{(r)} + i\gamma_i^{(r)})(\Omega_1 + \omega_j^{(J)} + i\gamma_j^{(J)})}P_J\\[0.15cm]
& \qquad\qquad \times \langle\langle G^{(r)},G^{(r)}|{\cal G}(T_2)|\psi_{j'}^{(J)},\psi_j^{(J)}\rangle\rangle \bigg)\ \tilde{{\cal E}}_{lo}^*(\omega_i^{(r)}-\omega_{lo})\tilde{{\cal E}}_3(\omega_i^{(r)}-\omega_3)\tilde{{\cal E}}_2(\omega_j^{(J)}-\omega_2)\tilde{{\cal E}}_1^*(\omega_j^{(J)}-\omega_1)\bigg]
\end{split}
\label{Sif}
\end{equation}
by Fourier transform with respect to $T_1$ and $T_3$ delays, where the Green's function during $T_2$ delay reads
\begin{equation}
\begin{split}
& \langle\langle \psi_i^{(r)},\psi_{i'}^{(r)}|{\cal G}(t)|\psi_j^{(J)},\psi_{j'}^{(J)}\rangle\rangle = \sum_{m,n=1}^{N+1}\sum_{k,l=1}^{N+1}\sum_{u=1}^{\text{dim}(L)} S_{\langle e_me_n,r\rangle;u}e^{\nu_u t}S_{u;\langle e_ke_l,J\rangle}^{-1} C_{e_k,j}^{(J),*}C_{e_l,j'}^{(J)}C_{e_m,i}^{(r)}C_{e_n,i'}^{(r),*}\\[0.15cm]
& \langle\langle G^{(r)},G^{(r)}|{\cal G}(t)|\psi_{j'}^{(J)},\psi_j^{(J)}\rangle\rangle = \frac{\omega_c}{Q}\sum_{q,P}\sum_{k,l=1}^{N+1}\sum_{u=1}^{\text{dim}(L)}\prod_{s=1}^N\langle\langle r_s,r_s|{\cal G}_s^{(gg)}(t)|q_s,q_s\rangle\rangle\ S_{\langle e_{N+1}e_{N+1},P\rangle;u}S_{u;\langle e_ke_l,J\rangle}^{-1}\\[0.15cm]
& \qquad\qquad\qquad\qquad\qquad\qquad\qquad\quad \times \left(\int_0^t\text{d}t' e^{\nu_u t'}\prod_{w=1}^N\langle\langle q_w,q_w|{\cal G}_w^{(qq)}(-t')|P_w,P_w\rangle\rangle\right)C_{e_k,j'}^{(J),*}C_{e_l,j}^{(J)}\\[0.15cm]
& {\cal G}_s^{(gg)}(t) = \frac{1}{2\bar{n}_{v_s}+1}\begin{pmatrix}
                                                    \bar{n}_{v_s}+1 & \bar{n}_{v_s}+1\\[0.15cm]
																										\bar{n}_{v_s} & \bar{n}_{v_s}
																									 \end{pmatrix}
																								+ \frac{e^{-\gamma_s(2\bar{n}_{v_s}+1)t}}{2\bar{n}_{v_s}+1}
																								    \begin{pmatrix}
																									   \bar{n}_{v_s} & -\bar{n}_{v_s}-1\\[0.15cm]
																										 -\bar{n}_{v_s} & \bar{n}_{v_s}+1
																								 		\end{pmatrix}
\end{split}
\label{Gt2}
\end{equation}
and $\nu_n$'s are the eigenvalues of Liouvillian $\hat{L}$ in Eq.(\ref{rho}), which governs the dynamics of the joint vibration/photon system during $T_2$ delay. $P_J$ denotes the statistical probability of system at $|G^{(J)},G^{(J)}\rangle\rangle$ prior to the pulse actions. For thermal equilibrium,
\begin{equation}
\begin{split}
P_J = 2^{-N}\prod_{s=1}^N\left[1 + (-1)^{\delta_{J_s,1}}\text{tanh}\left(\frac{\hbar v_s}{2k_B T}\right)\right]
\end{split}
\label{PJ}
\end{equation}

To reveal the effect of local fluctuations induced by the disorder, we first neglect the thermal excitations in  solvent before acting the pulses, for simplicity. This is to say that the molecules are at the vibrational ground state and the coordinates of solvent motion take $q_j=0;j=1,2,\cdots,N$, which gives $P_1=1;P_J=0\ (J=2,3,\cdots,2^N)$ prior to pulse action. Note the subscript $``1"$ denotes the $\{0_1,0_2,\cdots,0_N\}$ configuration of solvent coordinates. The thermal excitations in solvent will be taken into account later. After deducing the GSB contribution, Fig.\ref{2DIR} show the tomographies of 2D-IR signal $S_I(\Omega_3,T_2,\Omega_1)$ with different $T_2$ delays, for three W(CO)$_6$ molecules placed along the axis of an IR cavity. First of all it is shown that the line-broadening along anti-diagonal is larger than that along diagonal. This is reasonable because of the inhomogenous broadening attributed to the solvent-induced disorder effect. By introducing the $T_2$ delay, the cross peaks above the anti-diagonal show up and their intensities keep increasing. Those cross peaks with the position $\Omega_3\simeq 18$cm$^{-1}$ manifest the excitation transfer from either polariton (LP or UP) to dark states that nearly decouple with cavity modes. This is elucidated by the fixed probe frequency $\Omega_3$ without much shift as time propagates. The cross peak positioned at $\Omega_3\simeq -3.6$cm$^{-1}$ ($\Omega_3\simeq +3.6$cm$^{-1}$) provides the information about the excitation transfer from UP to LP (from LP to UP). For the polariton-dark-states transfer, Fig.\ref{2DIR} illustrates that the rate of UP $\rightarrow$ dark-states is higher than that of LP $\rightarrow$ dark-states. We attribute this to the faster decay of intermolecule coherence for UP than the one for LP, as elucidated in Fig.\ref{dm}(d) and \ref{dm}(e). Moreover, we can also observe in Fig.\ref{2DIR} that the excitation transfer between LP and UP is slower than the one between polaritons and dark modes, since the former occurs in $\sim 100$ps whereas the latter occurs in $\sim 30$ps. This is further supported by the population dynamics depicted in Fig.\ref{dm}(g) and \ref{dm}(h). 

\begin{figure}
 \captionsetup{justification=raggedright,singlelinecheck=false}
 \centering
   \includegraphics[scale=0.35]{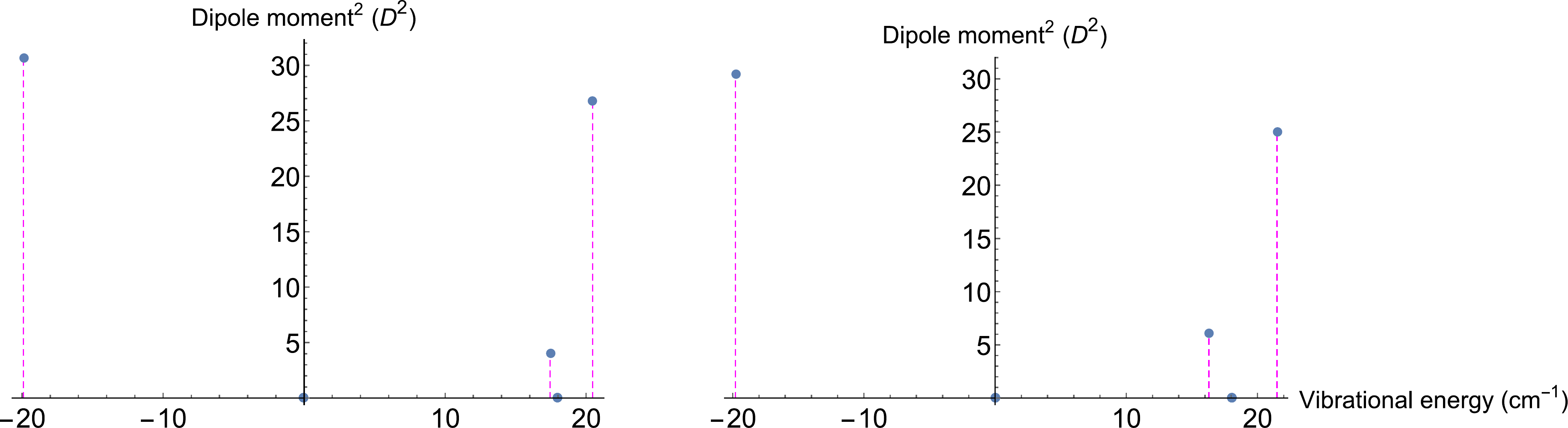}
\caption{Dipole moment strength varies with energy levels of the joint vibration/photon system, for 4137 W(CO)$_6$ molecules placed in an IR cavity with a linear fashion along the cavity axis. (Left) 30 and (right) 130 out of 4137 W(CO)$_6$ molecules are large-detuned to cavity photons. $g_j\sqrt{N}=19$cm$^{-1}$ and $N=4137$ \cite{Joel_2017}. Other molecular parameters are the same as Fig.\ref{dm}.}
\label{dipole}
\end{figure}

With the deduction of GSB contribution, Fig.\ref{2DIRf} shows the 2D-IR signal $S_I(\Omega_3,T_2,\Omega_1)$ with different $T_2$ delays, by considering the thermal excitations on solvent degrees of freedoms whereas the molecules are at vibrational ground state prior to the pulse actions. Compared to Fig.\ref{2DIR}, we observe the cross peaks both above and below the anti-diagonal line. In addition to the information given by Fig.\ref{2DIR}, the excitation transfer from dark states to LP \& UP is evident by the cross peaks below the anti-diagonal, when dark states are excited. The dark states $\rightarrow$ polaritons transfer is faster than the one between LP and UP, which is manifested by the polariton dynamics shown in Fig.\ref{dm}(i). So far, the analysis based on Fig.\ref{2DIR} and Fig.\ref{2DIRf} clearly elucidates how the cooperativity in the joint vibration/photon system is eroded by the local fluctuations (i.e., disorder), associated with the presence of extra cross peaks other than those characterizing the polariton modes. Hence this enables us to explain and gain more understanding for recent experiments \cite{Xiang_PNAS2018} where the coupling between bright polariton and dark modes was demonstrated. However, the position of dark states relative to polaritons in 2D-IR spectra predicted by our work shows the deviation from the experiments \cite{Xiang_PNAS2018}. This is due to the fact of the much lower amount of molecules considered in our calculations (for $N=3$, $g\sqrt{N}\simeq 3.6$cm$^{-1}\ll \delta\omega$) than the case in real experiments which placed $\sim 4000$ W(CO)$_6$ molecules in an IR cavity. To support this, we further calculate the distribution of dipole moments for an ensemble with 4137 W(CO)$_6$ molecules included, as depicted in Fig.\ref{dipole}, where (left) and (right) correspond to the cases of 30 and 130 out of 4137 molecules being large-detuned to photons under the influence of solvent-induced disorder. The result manifests the extra peak sandwiched in between the two polariton branches. This is attributed to the dark modes since the associated frequency shift of a few cm$^{-1}$'s (compare Fig.\ref{dipole} (left) and 6(right)) is much smaller than the Rabi splitting $\simeq 2g\sqrt{N}\simeq 40$cm$^{-1}$ between the two polariton branches. These, combined with Fig.\ref{2DIR} and Fig.\ref{2DIRf}, thereby captures the feature in consistence with the experimental results \cite{Xiang_PNAS2018}. The full simulation of the dynamics in terms of QME or QSLE given by Eq.(\ref{qme}) for number of molecules demands a heavy effort of computation, which goes beyond the scope of this paper and would be performed in the future research.
 
\section{Conclusion and remarks}
We studied the collective properties of the vibrational polaritons, by developing time-resolved photoluminescence and two-dimensional infrared spectroscopies incorporating the disorder induced by solvent motion. Our results demonstrated that vibrational excitations become localized, associated with the cross peaks positioned at fixed probe frequency in 2D-IR spectra, when the cooperativity between molecules is diluted by solvent-induced local noise. Understanding the cooperative nature of polaritons is significant for the community to gain more details about the dark states weakly interacting with cavity fields. The information about such dark states has elucidated the intimated connection to the design of vibrational-polariton photonic devices in mid-IR regime.

As inspired by recent advance in 2D-IR of vibrational polaritons \cite{Xiang_PNAS2018,Joel_2017,Owrutsky_NatCom2016}, our work can offer new insight for understanding the mechanism of polariton-dark-states couplings evident by the measurement in Ref.\cite{Xiang_PNAS2018}. The present work can be extended to the case of number of molecules incorporating the solvent motion described by continuous coordinates. This shows the perspective for considerably improving the theories in quantitative agreement with experiments. The formalisms developed in present paper can be further generalized to the exciton process in vis/UV regime, by integrating the nuclei motions. Taking the advantage of the lower cost compared to the full simulation of nuclei wavepacket \cite{Agarwalla_JCP2015,Sanda_JPCB2008}, this would pave an alternative road for studying the cavity-controlled charge transfer and reaction kinetics.

\end{document}